\documentclass[a4,12pt]{article}

\RequirePackage[OT1]{fontenc}

\usepackage{amsmath, amsthm, amssymb}
\newtheorem{proposition}{Proposition}

\usepackage{amsmath}
\usepackage[english]{babel}
\usepackage[ansinew]{inputenc}
\usepackage{epsfig}
\usepackage{color,graphicx}
\usepackage{wrapfig}
\usepackage{amssymb,amsthm,amsmath}
\usepackage{dsfont}
\usepackage{natbib}
\usepackage{abstract}

\usepackage{enumerate}

\renewcommand{\abstractname}{Summary}

\newtheorem*{definition*}{Definition}
\newtheorem*{thm*}{Theorem}
\newtheorem*{proposition*}{Proposition}

\def\ds{\displaystyle}

\begin{document}

\title{\bf Barker's algorithm for Bayesian inference with intractable likelihoods}
\author{
Flavio B. Gon\c calves$^{a}$,
Krzysztof {\L}atuszy\'nski$^{b,c}$,
Gareth O. Roberts$^{b}$
}
\date{}

\maketitle

\begin{center}
{\footnotesize $^a$
Departamento de Estat\'istica, Universidade Federal de Minas Gerais
\\
$^b$
Department of Statistics, University of Warwick \\
$^c$ Alan Turing Institute}
\end{center}

\footnotetext[1]{Address: Av. Ant�nio Carlos, 6627 - DEST/ICEx/UFMG - Belo Horizonte, Minas Gerais, 31270-901, Brazil. E-mail: fbgoncalves@est.ufmg.br}


\renewcommand{\abstractname}{Abstract}
\begin{abstract}
In this expository paper we abstract and describe a simple MCMC scheme for sampling from intractable target densities. The approach has been introduced in \cite{1707.00332} in the specific context of jump-diffusions, and is based on the Barker's algorithm paired with a simple Bernoulli factory type scheme, the so called \emph{2-coin algorithm}. In many settings it is an alternative to standard Metropolis-Hastings pseudo-marginal method for simulating from intractable target densities. Although Barker's is well-known to be slightly less efficient than Metropolis-Hastings, the key advantage of our approach is that it allows to implement the ``marginal Barker's'' instead of the extended state space pseudo-marginal Metropolis-Hastings, owing to the special form of the accept/reject probability. We shall illustrate our methodology in the context of Bayesian inference for discretely observed Wright-Fisher family of diffusions.


\end{abstract}

\section{Introduction}

Modern data science is awash with problems with intractable likelihoods, i.e. problems in which pointwise evaluation of the likelihood function is either impossible or extremely computationally expensive. Intractability can be caused by data missingness, model complexity, or the sheer size of the data set. Within this context, Bayesian inference is
particularly challenging, as its algorithmic workhorse, MCMC,  requires large numbers (typically many thousands) of likelihood evaluations, and therefore runs the risk of being
prohibitively slow.

\def\state{{\mathcal X}}
We place our problem in the following generic context. Suppose that
\mbox{$\pi(\theta) =\pi (\theta | y)$} is a target (posterior) density
(with respect to some dominating measure $\nu $) of parameter $\theta $ on state space $\state $ given data set $y$.
Considering the problem of designing an MCMC algorithm that targets
$\pi,$ we shall use  $q(\theta , \cdot )$ to denote the density (also with respect to $\nu $) of the proposed transition from $\theta $. The standard Metropolis-Hastings algorithm proposes a move from
$\theta $ to $\phi$ accepting with probability
\begin{equation}
\label{e:MHaccept}
\alpha _{MH} (\theta , \phi ) =
1 \wedge {  \pi (\phi ) q(\phi ,\theta ) \
\over
\pi (\theta ) q(\theta ,\phi ) }
\end{equation}
requiring at least a function evaluation of $\pi (\phi )$.

Motivated by this, in recent years there has been increased interesting {\em retrospective} simulation techniques which  attempt to simulate from an event of probability (\ref{e:MHaccept})
directly, and without recourse to calculating the probability itself,
see e.g.
\cite{bpr07,bpr06a,papaspiliopoulos2008retrospective}.
These methods rely heavily on being able to simulate from
events such as a duplicated data set, $y'$ say, conditional on the parameter $\phi $ which might have probability (proportional to) $\pi (\phi )$.

However in the current context, even if retrospective simulation of
events of probability $\pi (\theta )$ and $\pi (\phi )$ is possible
and efficient, that does not directly lead to a solution to the
problem of simulation from an event of probability~(\ref{e:MHaccept})
which is a nonlinear expression of $\pi (\theta )$ and $\pi (\phi
)$. This problem falls into the category of the classical
computational probability problem known as the {\em Bernoulli factory
  problem}: given the probability $p$ which we cannot evaluate but
where events of this probability can be simulated, how can one
simulate from an event of probability $f(p)$?
(c.f. \cite{von1951various,keane1994bernoulli}). In the context
of~(\ref{e:MHaccept}), the function $f$ takes the form $f(p_1, p_2) = 1 \wedge { c_1 p_1
\over
c_2 p_2 }$  for suitable constants $c_1$ and $c_2$, and in many
modelling contexts of interest can be equivalently rewritten as
\begin{equation}
\label{e:MHf}
f(p_3) = 1\wedge c_3p_3.
\end{equation}
While substantial progress has been made on efficient solutions to the
Bernoulli factory problem in recent years
(e.g. \cite{nacu2005fast,p2p} and in more specialised settings \cite{flegal2010exact,herbei_berliner,MR3506427,huber2015optimal}), it is known that in general
there does not exist a solution when $f$ takes the form given in
(\ref{e:MHf}), c.f. \cite{asmussen1992stationarity,p2p}. Fortunately, the flexibility of the Metropolis-Hastings algorithm allows us to circumvent this difficulty.

In this paper we will abstract and describe a  framework for the
implementation of MCMC with intractable likelihoods by using an
alternative acceptance probability to that in  (\ref{e:MHaccept});
namely that introduced in \citet{Barker}. Although less efficient than
(\ref{e:MHaccept}), the Barker acceptance probability can be simulated
using an efficient and simple Bernoulli factory type algorithm
we will describe,  called the {\em 2-coin algorithm}. The approach has
been first used in the complex setting of exact fully Bayesian inference for
stochastic differential equations with jumps (\cite{1707.00332}) and then
also for diffusions with
switching regimes (\cite{LaPaRo_MarkovSwitching}) and diffusion driven Cox processes \citep{GRL2}.
Here, for expository purposes we will illustrate our algorithm with one
toy example and then move on to describe a more realistic example of Bayesian inference for discretely
observed Wright-Fisher diffusions for which the approach of
\cite{bpr06a} is not applicable.

\section{Barker's alternative to Metropolis-Hastings}

Peskun \citep{Peskun} demonstrated that given a proposal kernel $q$,
there are many different choices of acceptance probability
$\alpha(\theta, \phi)$ which create a Markov chain with stationary
distribution $\pi $, going on to show that the Metropolis-Hastings
option $\alpha_{MH}(\theta, \phi)$ in~(\ref{e:MHaccept}) maximises the
acceptance probability for all possible transitions $\theta
\rightarrow \phi $. This in turn implies (via the celebrated Peskun
ordering) that the Metropolis-Hastings choice of acceptance
probability minimises the asymptotic Monte Carlo variance for
estimating integrals of functions in $L^2(\pi)$  (see also
\cite{mira1999ordering, mira2001ordering}; here $L^2(\pi)$ denotes
the Hilbert space of functions square integrable with respect to $\pi$,
for background on functional analytic perspective on Markov
chains we refer to \cite{roberts1997geometric} and references therein). However, there are many other acceptance probability solutions which can be shown to be almost as good. We will focus on one introduced in \citet{Barker}:
\begin{equation}
\label{e:Barkacc}
\alpha _B (\theta , \phi ) =
 { \pi (\phi ) q(\phi , \theta ) \over \pi (\theta ) q(\theta ,\phi)  + \pi (\phi ) q(\phi , \theta )}\ .
\end{equation}
It is straightforward to verify that
$$
{\alpha _{MH} (\theta , \phi ) \over 2} \le \alpha _{B} (\theta , \phi ) \le \alpha _{MH} (\theta , \phi ),
$$
which indicates that the two algorithms have similar performance. More
precisely, in terms of comparing the CLT and its asymptotic variance, the following
result holds, which demonstrates that roughly speaking Barker's method is  at worst half as good as Metropolis-Hastings.

  \begin{proposition}[\cite{latuszynski2011clts}, Theorem 4(ii)] Let
    $f \in L^2 (\pi )$ and denote the iid Monte Carlo variance by
    $\sigma^2_{\pi}:=\textrm{Var}_{\pi}(f)$. If a square root central limit
    theorem holds for $f$ and the Metropolis-Hastings chain with the
    CLT assymptotic variance $\sigma^2_{MH}$, i.e.
$$
{\sum_{i=1}^N f(\theta _i) - \pi (f) \over \sqrt{N}} \Rightarrow N(0, \sigma_{MH}^2),
$$
then a corresponding CLT holds for $f$ and the Barker chain with CLT
asymptotic variance $\sigma _B^2$ satisfying
$$
\sigma_{MH}^2  \le  \sigma_{B}^2 \le 2 \sigma_{MH}^2 + \sigma^2_{\pi}. $$
\end{proposition}

On the other hand, when we can simulate efficiently from events proportional to $\pi (\theta )$, the following {\em 2-coin algorithm} gives a simple way of  simulating from an event of probability $\alpha _B(\theta , \phi )$, which we write as $\ds \frac{c_1p_1}{c_1p_1+c_2p_2}$:\\
\\
\begin{tabular}[!]{|l|}
\hline
\parbox[!]{12cm}{
\smallskip
\underline{The {\em 2-coin algorithm} for sampling Barker's acceptance probability:}
{\footnotesize
\begin{enumerate}\label{2ca}
\setlength\itemsep{-0.1em}
  \item Sample $\ds C_1\sim Ber\left(\frac{c_1}{c_1+c_2}\right)$;
  \item if $C_1=1$, sample $C_2\sim Ber(p_1)$;
  \begin{itemize}
    \item if $C_2=1$, output 1;
    \item if $C_2=0$, go back to 1;
  \end{itemize}
  \item if $C_1=0$, sample $C_2\sim Ber(p_2)$;
  \begin{itemize}
    \item if $C_2=1$, output 0;
    \item if $C_2=0$, go back to 1.
    \end{itemize}
\end{enumerate}}
}\\  \hline
\end{tabular}

\medskip
Elementary conditional probability calculations verify that the above
{\em 2-coin algorithm} outputs $1$ with probability $
\frac{c_1p_1}{c_1p_1+c_2p_2}$ and $0$ with probability $
\frac{c_2p_2}{c_1p_1+c_2p_2}.$ Furthermore, the number of loops needed
until the algorithm stops is distributed as $\textrm{Geom}(\frac{c_1p_1+c_2p_2}{c_1
+ c_2})$ and hence the mean execution time is proportional to  $\frac{c_1
+ c_2}{c_1p_1+c_2p_2}.$

The approach described in this paper can be seen as an alternative to
pseudo-marginal MCMC \citep{beaumont,ARPM} in settings where simulation of an event of probability proportional to $\pi(\theta)$ is possible, hence in particular when there is an unbiased positive and bounded estimator of $\pi(\theta)$ (c.f. Lemma 2.1 and Theorem 2.7 of \cite{p2p} and also \cite{jacob2015nonnegative} for a related discussion). While both methods have
their advantages, we see that the Barker approach advocated here has
the clear distinction of constructing the marginal algorithm, i.e. a
Markov chain on  the state space $\state$ without any need for
additional auxiliary variables, which slow down
pseudo-marginal methods in terms of the CLT asymptotic variance
(see \cite{MR3313762}, Theorem 7). This paper will not attempt a systematic numerical comparison between these approaches.
\medskip

\section{A simple example}\label{sec:toy}

\def\expect{{\bf E}}
To illustrate how an intractable posterior density $\pi$ may be
written as a product of a known function and an unknown probability from
which one can simulate, so that it can be readily sampled via the above
Barker's algorithm with 2-coin acceptance, consider the following simple (toy) example. Suppose that
(perhaps due to missing data) we can write the posterior distribution $\pi $
as
$$
\pi (\theta ) = \expect_{\eta \sim h}[ \pi (\theta \mid \eta )   ]
$$
that is, we only have an explicit expression for $\pi $ as a mixture  of conditional densities $\pi (\cdot \mid \eta ) $ with known mixing measure $h$.

Now, we shall be even more explicit, and assume that $\theta \mid \eta \sim $Poisson$(\eta )$ so that
$$
\pi (\theta \mid \eta ) ={e^{-\eta } \eta ^{\theta} \over \theta ! } \le {e^{-\theta } \theta ^{\theta} \over \theta ! } := d(\theta ).
$$

Writing $\pi(\theta) = d(\theta) \frac{\pi(\theta)}{d(\theta)}$ yields
the desired form since events of probability $p(\theta ) :=\pi (\theta )/d(\theta )$ are easy
to simulate as follows. First simulate $\eta \sim h$ and an
independent standard uniform random variable $U$. Then the event $\{ U\le \pi (\theta  \mid \eta )/d(\theta )\} $ is easily seen to have the desired probability.

Now consider a Barker algorithm which proposes from a symmetric random walk move (ie $q(\theta , \phi ) = q(\phi , \theta )$). We are thus required to accept the proposed move with probability
$$
\alpha _B(\theta , \phi ) = {d(\phi ) p(\phi ) \over  d(\phi ) p(\phi )  + d(\theta ) p(\theta ) }
$$
which we can do using the 2-coin algorithm.

\begin{figure}[h]
	\centering
	\includegraphics[scale=0.35]{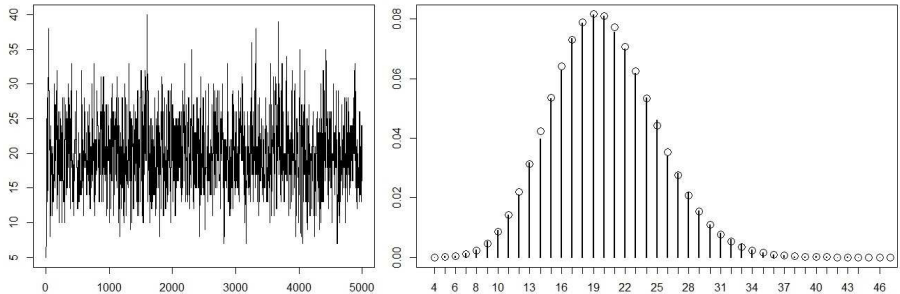}
	\caption{Trace plot for the first 5 thousand iterations and
          empirical (lines) and exact (circles) distribution
          $\pi(\theta)$. The estimated mean and variance are 20.012
          and 23.989.}\label{f2}
\end{figure}

We implement this example with $\eta\sim Gamma(100,5)$ and a uniform proposal distribution on $\{\theta-10,\theta-9,\ldots,\theta-1,\theta+1,\ldots,\theta+9,\theta+10\}$, which leads to a 0.367 acceptance rate. The exact distribution $\pi(\theta)$ is Negative Binomial with mean 20 and variance 24. We run the chain for $2.10^6$ iterations. Results are reported in Figure \ref{f2}. The average number of loops in the 2-coin algorithm is 4.7.

\section{Barker's MCMC for exact inference in the Wright-Fisher family of diffusions}\label{secBfD}

In this section we sketch the MCMC methodology for exact Bayesian inference for diffusions introduced in \cite{bpr06a} and further developed in \cite{sermai}. We then consider inference for the Wright-Fisher diffusion and explain why it falls outside the scope of this methodology. Next, for the Wright-Fisher family of diffusions, we present how to design the Barker's based MCMC algorithm, originally proposed in \citet{1707.00332} in the context of jump-diffusions. The algorithm performs exact Bayesian inference by sampling from a Markov chain that has the exact posterior distribution of the parameters and missing paths of the diffusion as its invariant distribution.

\subsection{Exact Bayesian inference for discretely observed diffusions}
The model considered in this section is that of a stochastic process $Y:=\{Y_s:0\leq s\leq t\}$ that solves the following SDE parametrised by $\theta$:
\begin{equation}\label{e1}
\ds dY_s=b(Y_s;\theta)ds+\sigma(Y_s;\theta)dW_s,\;\;Y_0=y_0,
\end{equation}
where $W_s$ is a Brownian motion and the drift coefficient $b:\mathds{R}\rightarrow\mathds{R}$ and the diffusion coefficient $\sigma:\mathds{R}\rightarrow\mathds{R}$ are such that there exists a unique weak solution to \eqref{e1} \citep[for background on diffusions see e.g.][in particular chapter 4 for stability conditions]{KP95}. We consider only univariate and time-homogeneous processes.

Suppose that the data \[Y_{obs} =(y_0,y_1,\ldots,y_n)\] comes from $Y$ observed at times $0=t_0,\;t_1,\ldots,\;t_{n-1},\;t_n=T$.

The interest is in the parameter $\theta$ and we work in the Bayesian setting, assuming a prior $\pi(\theta)$ on $\theta$. Hence, if we denote the transition densities of $Y$ as
\begin{equation} \label{trans_dens}
p(\theta, y_{i-1}, y_{i}) := \mathbb{P}_{\theta}\big(Y_{t_{i}} \in d y_{i} | Y_{t_{i-1}}= y_{i-1} \big)/ d y_{i}, \end{equation}
the posterior distribution of $\theta$ given data $Y_{obs}$ becomes
\begin{equation}\label{posterior}
\pi(\theta| Y_{obs}) \propto \pi(\theta)\prod_{i=1}^n p(\theta, y_{i-1}, y_i).
\end{equation}
Unfortunately the diffusion transition density in \eqref{trans_dens} is not available in closed form except for the constant coefficients case, and a few other very special diffusion families. Consequently, standard inference about $\theta$ relies on approximations, such as Euler approximation of the diffusion dynamics, or Monte Carlo approximations of transition densities. We refer to \cite{bpr06a, sermai, 1707.00332} for discussions of the approximation based inference for diffusions, and in particular the difficult to quantify bias that it introduces, and how the computational cost scales as the desired inferential error decreases.

An alternative to the approximation based inference is the exact inference for diffusions introduced in \cite{bpr06a}. The approach avoids any discretisation error and allows to design MCMC algorithms that target the exact posterior \eqref{posterior} on an extended state space via introducing an auxiliary variable $Y_{mis}$ that represents the missing continuous paths of $Y$ between discrete observations $Y_{obs}.$ The first approximation of how such an MCMC algorithm would be designed, is to conceptualise a Gibbs sampling alternating the following two steps:
\begin{eqnarray}\label{gibbs1_path}
Y_{mis} & \sim &\mathcal{L}\big(Y_{mis} | Y_{obs}, \theta \big); \\ \label{gibbs1_para}
\theta & \sim & \mathcal{L} \big( \theta | Y_{obs}, Y_{mis} \big).
\end{eqnarray}
The practical execution of the above Gibbs algorithm, in particular of step~\eqref{gibbs1_path}, is based on exact sampling of diffusion bridges developed in \cite{bpr06b, bpr07}, termed Exact Algorithm (EA). However, for EA to be applicable, we need to work a bit more. Exact Algorithm is a rejection sampler on the diffusion path space. It proposes a diffusion bridge between two data points from a driftless diffusion (i.e. one with $b \equiv 0$) and accepts or rejects this proposal based on the Radon-Nikodym derivative given by the Girsanov theorem. In \eqref{e1} the diffusion coefficient $\sigma$ depends on $\theta$ and this presents a problem as there is a perfect correlation between the diffusion path (precisely speaking, its quadratic variation) and the diffusion coefficient, so the Gibbs sampler above would not mix at all. Secondly, the driftless diffusion bridge proposals for $Y_{mis}$ in the EA need to be simulable, which in practice means that
the diffusion coefficient in \eqref{e1} can not depend on $Y_s$.

Hence, in order to perform MCMC inference, the diffusion \eqref{e1} is transformed into a stochastic differential equation with unit diffusion coefficient using Lamperti transformation.
This allows us to obtain a parameter-free dominating measure to write the likelihood function of a complete diffusion path, and in particular to design diffusion Bridge proposals based on the standard Brownian motion. The Lamperti transform
\begin{equation}
\ds X_s=\eta(Y_s;\theta)=\int_{v}^{Y_s}\frac{1}{\sigma(u;\theta)}du,
\end{equation}
 where $v$ is some arbitrary element of the state space of $Y$, implies that $X$ is a diffusion solving the SDE
\begin{equation}\label{e2}
\ds dX_s=\alpha(X_s;\theta)ds+dW_s,\qquad X_0=x_0(\theta)=\eta(y_0,\theta),
\end{equation}
where
\begin{equation} \label{def_alpha} \ds \alpha(u;\theta)=\frac{b(\eta^{-1}(u;\theta);\theta)}{\sigma(\eta^{-1}(u;\theta);\theta)}-\frac{\sigma'(\eta^{-1}(u;\theta);\theta)}{2}.\end{equation}
Note that $X_s$ is a function of $\theta$, in
particular the observed data becomes
 $$(x_0(\theta),\ldots,x_n(\theta))=(\eta(y_0;\theta),\ldots,\eta(y_n;\theta)).$$
The second transformation, now only for the bridges of $X$ (between
the observations), is given by
\begin{eqnarray}\label{inverse_psi} \dot{X}_s & = & \varphi_{\theta}^{-1}(X_s)
                                                    \nonumber
  \\ \label{inverse_psi}  & := & X_s -
\left(1-\frac{s-t_{i-1}}{t_i-t_{i-1}}\right)x_{i-1}(\theta)-
\left(\frac{s-t_{i-1}}{t_i-t_{i-1}}\right)x_{i}(\theta), \end{eqnarray}
for $s\in(t_{i-1},t_i)$.
This transformation guarantees that the measure of each bridge
$\dot{X}_s$ is dominated by the measure of a standard Brownian bridge
of same time length. We denote the transformation in \eqref{inverse_psi}
as $\varphi_{\theta}^{-1}$, so that we have
the handy notation  $\varphi_{\theta}(\dot{X}_s)$ for its inverse.

By $\dot{X}_{mis}$ denote the bridges of $\dot{X}.$
Defining $Y_{com}=\{Y_{obs},\dot{X}_{mis}\}$, where $Y_{obs}=(y_0,y_1,\ldots,y_n)$ and $X_{mis}$ are all the bridges of $\dot{X}$, Lemma 2 from \citet{bpr06a} gives the likelihood $L(\theta;Y_{com})$ of a complete diffusion path in $[0,T]$ and the joint posterior density of $(\theta,X_{mis})$ satisfies $\ds    \pi(\theta,X_{mis}|Y_{obs})\propto L(\theta;Y_{com})\pi(\theta)$, which gives us the full conditional densities for the Gibbs sampler, as detailed in Sections \ref{sampmp} and \ref{samppar}.

Exact methodologies for inference in diffusion processes have been
firstly proposed in \citet{bpr06a} and are based on the Exact
Algorithm (EA), which simulates a class of diffusion processes exactly
via rejection sampling. However, the feasibility of all those
methodologies rely on the assumption that a given function of the diffusion components is
bounded below (see condition \eqref{cond_G_integrant} below), while their efficiency depends on the tightness of this bound. Although broad, the class of processes that satisfy this assumption exclude some appealing processes, such as the Cox-Ingersoll-Ross model and Wright-Fisher family of diffusions \citep{JaS}. The methodology we present in this section is general enough not to require such assumption and, therefore, can be applied to models such as the ones just mentioned. We present an example with a model from the Wright-Fisher family.

As mentioned, the existing exact
inference methodologies rely heavily on the assumption that a certain
function is bounded below. The function in question is
$\ds\left(\alpha^2+\alpha'\right)(u;\theta),$ where $\alpha$ is defined in \eqref{def_alpha}, and the derivative is
in the space variable $u$. In case of one dimensional real valued
diffusions, the precise condition reads
\begin{equation} \label{cond_G_integrant}\ds \inf_{u \in
    \mathbb{R}}\left\{\left(\alpha^2+\alpha'\right)(u;\theta) \right\}
  \; \geq \; a(\theta) \;>\; -\infty. \end{equation}
 Hence the function is required to be uniformly bounded below in the state space of $X$, for all $\theta$ in the parametric space. The methodology presented here, however, does not require this boundedness assumption, as we make it clear further ahead in this Section.

\subsection{Sampling the missing paths}\label{sampmp}

The transformed missing bridges $X_{mis}$ are sampled via Barker's with
standard Brownian bridge proposals. Given that the bridges are
conditionally independent, due to the Markov property, the Barker's
step for each bridge may be performed in parallel.  Our description shall
focus on the update step of a single interval: $(t_{i-1},t_i)$.

The Barker's acceptance probability $\alpha_X$ of a proposed bridge
$\dot{B}$ given a current bridge $\dot{X}$ in $(t_{i-1},t_i)$ is obtained using the measure of a standard Brownian bridge as the dominating measure and is given by
$$
\alpha _X  = \alpha _X(\dot{X}, \dot{B}) =  \frac{G(\varphi_{\theta
  }(\dot{B}); \theta)}{ G(\varphi_{\theta }(\dot{B}) ; \theta) + G(\varphi_{\theta }(\dot{X}) ; \theta)},
$$
where $G$ is derived from Girsanov's formula (see for example \cite{sermai} in a similar context): for an arbitrary path $Y$
$$
G(Y ; \theta) =
\exp\left\{-\int_{t_{i-1}}^{t_i}\left(\frac{\alpha^2+\alpha'}{2}\right)(Y_s ; \theta)ds\right\} \ .
$$
Now we intend to follow the strategy of conditioning by an auxiliary
variable adopted in the simple example of
Section \ref{sec:toy}. However, to do this we require an upper bound on $G$.
For some diffusion models the functional form of $\alpha^2+\alpha'$ is bounded below which leads to an upper bound for $G$. However for many models (including the Wright-Fisher model we shall go on to consider) we will need to have additional information about the sample path $Y$ to provide the necessary bounds. To this end, we adopt the {\em layered Bownian bridge} construction
of \cite{bpr06a, bpr07}. Giving a detailed description of this construction
is beyond the scope of this paper though the complete details can be
found in \cite{bpr07,1707.00332}. The important feature of this
construction for our purposes is that the layer of a path, in
particular of the proposal path  $\dot{B}$, can be directly simulated,
and leads to upper and lower bounds on its potential trajectories that
are of the form $L < \inf_{s \in (t_{i-1}, t_i)}\dot{B}_s < \sup_{s \in (t_{i-1}, t_i)}\dot{B}_s < U$ for
some $L, U \in \mathbb{R}$. This in turn allows us to produce a local
lower bound on of $\alpha^2+\alpha'$ on the compact set $[L, U]$ which
apply for any path consistent with that layer. For trajectory $X$ we
shall call this lower bound $2a_i(X; s,\theta )$, so that
\begin{equation} \label{ai}
\frac{(\alpha^2+\alpha')(\varphi _{\theta } (\dot{X}_s)}{2} \ge a_i(X; s,\theta )
\ ,
\end{equation}
for all paths $X$ consistent with the  layer of $X$. Note that $a_i$ may not depend upon $s$ as we are here only considering $s\in [t_{i-1}, t_i]$ although in the next section we shall need to consider dependence on $s$ as $X$ will have different layers in different intervals. Furthermore, tighter bounds may be obtained if the layers are obtained for a standard bridge and then transformed to the original one, making the bounds a linear function of $s$.
Thus we shall write
\begin{eqnarray}
\alpha_X & = & \frac{s_Bp_B}{s_Bp_B+s_Xp_X},\;\;  \textrm{where}  \\
\label{apbak4}
 s_{X}&=&\exp\left\{-\mathcal{I}_{a_i}(\theta)\right\},\\ \label{px} \ds
             p_{X} & = & \exp\left\{-\int_{t_{i-1}}^{t_i}\left(\frac{\alpha^2+\alpha'}{2}\right)(\varphi_{\theta}(\dot{X}_s);\theta)-a_i(X;s,\theta)ds\right\}, \nonumber
\\
\ds \mathcal{I}_{a_i}(\theta)& =&
                                     \int_{t_{i-1}}^{t_i}a_i(X;s,\theta)ds,
                                     \quad \textrm{moreover,} \\
  \ds a_i(X;s,\theta) & \leq &
                             \left(\frac{\alpha^2+\alpha'}{2}\right)(\varphi_{\theta}(\dot{X}_s);\theta),\qquad s\in[t_{i-1},t_i].\end{eqnarray}
Consequently, $s_B$ and $s_X$ are known positive numbers and $p_B$ and
$p_X$ are unknown probabilities. However, coins with
probabilities  $p_B$ and
$p_X$ can be simulated using an algorithm called the Poisson
coin. We refer to \cite{bpr06b, bpr07,1707.00332} for a detailed
construction and use of the Poisson coin, the brief description is as
follows.

Note that by \eqref{ai} the integrand in \eqref{px} is strictly
positive and the layered Brownian bridge construction will also yield
a bound $r_i(X, \theta) \in \mathbb{R}_+$, valid for any path
consistent with the simulated layer, and such that
\begin{equation}
r_i(X, \theta) \geq \sup_{s \in (t_{i-1}, t_i)} \left\{
\left(\frac{\alpha^2+\alpha'}{2}\right)(\varphi_{\theta}(\dot{X}_s);\theta)-a_i(X;s,\theta)
\right\}.
\end{equation}
Now let $\Phi$ be a Poisson process with intensity $r_i(X, \theta)$ on
$[t_{i-1}, t_i] \times [0,1]$. Its realisation is a collection of points
$\{\phi_k, \chi_k\}_{k=1}^{\kappa}$ on $[t_{i-1}, t_i] \times [0,1],$ where $\kappa \sim
\textrm{Poiss}\big[(t_i - t_{i-1}) r_i(X, \theta)\big].$ If $N$ is the
number of points below the graph
\begin{equation*}
s \longrightarrow \frac{\left(\frac{\alpha^2+\alpha'}{2}\right)(\varphi_{\theta}(\dot{X}_s);\theta)-a_i(X;s,\theta)}{r_i(X, \theta)},
\end{equation*}
then
\begin{equation*}
\mathbb{P}[N=0|X] = p_X.
\end{equation*}
Sampling the Poisson process $\Phi$ is standard and verifying if its
points are below or above the graph requires revealing $\dot{X}_s$ at
a finite collection of timepoints $\phi_1, \dots, \phi_{\kappa}$,
which in the case of Brownian bridge proposals is also routine.

Thus all
the steps of the 2-coin algorithm can be performed in order to sample
an event of probability $\alpha_X$ and accept or reject the proposed Brownian
\mbox{bridge $\dot{B}$} in the missing path update step of the Gibbs sampler.


\subsection{Sampling the parameters}\label{samppar}

The parameter vector $\theta$ is also sampled by a Barker's step. The
proposal distribution is a symmetric random walk, but not necessarily
Gaussian. The Barker's acceptance probability $\alpha_{\theta}$ for a proposal
$\theta^*$, given a current value $\theta$, is obtained using the dominating measure from the prior density $\pi(\theta)$ and is given by
\begin{eqnarray*}
\ds \alpha_{\theta} & = &
                          \frac{s_{\theta^*}p_{\theta^*}}{s_{\theta^*}p_{\theta^*}+s_{\theta}p_{\theta}},\;\;
\textrm{where} \\
\ds s_{\theta} & = &
                     \exp\left\{A(x_n(\theta);\theta)-A(x_0(\theta);\theta)-\sum_{i=1}^n\mathcal{I}_{a_i}(\theta)\right\}\pi(\theta),\\
p_{\theta} & = &
                 \exp\left\{- \sum_{i=1}^n \int_{t_{i-1}}^{t_i}
                 \left(\frac{\alpha^2+\alpha'}{2}\right)(\varphi_{\theta}(\dot{X}_s);\theta)-a_i(X;s,\theta)ds\right\},\\
A(u;\theta) & = &
                 \int _0^u\alpha(y;\theta)dy.
\end{eqnarray*}

Once again, the acceptance decision of probability $\alpha_{\theta}$ can be
performed using the 2-coin algorithm and the Poisson coin in a manner
analogous to that described
in Section \ref{sampmp}.
The efficiency of the proposed methodology relies heavily on the probability of the two second coins which, in turn, relies on the lower bounds $a_i(X;s,\theta)$.
A detailed description of how to obtain efficient lower bounds $a_i(X;s,\theta)$ can be found in \citet{1707.00332}.

Finally, note that our MCMC steps do not require function $(\alpha^2+\alpha')$ to be bounded below. The form of the Barker's acceptance probability and the dynamics of the 2-coin algorithm requires only that bounds for the diffusion path are obtained. More specifically, the bounds for the Brownian bridge construction are used to obtain a lower bound for $(\alpha^2+\alpha')$  which allows us to write the Barker's acceptance probability in the 2-coin algorithm form.

\subsection{Example}

The Wright-Fisher family of diffusions
\begin{equation*}
\ds dY_s=\beta(Y_s)ds+\sqrt{Y_s(1-Y_s)}dW_s
\end{equation*}
is widely used in statistical applications, especially in genetics,
see \cite{schraiber2013analysis} and references therein. It
is an example for which after applying the lamperti transformation the
function $(\alpha^2+\alpha')$ of the resulting drift $\alpha$ is not bounded below. \citet{JaS} propose algorithms to perform exact simulation of processes in that family but, although this could potentially be used to developed exact inference methodology, the authors do not pursue this direction in their paper.

We apply the Barker's methodology presented above to perform exact
inference about drift parameters for the neutral Wright-Fisher diffusion with mutation which
admits the following parametric SDE:
\begin{equation}\label{WFd}
\ds
dY_s=\frac{1}{2}\left(\theta_1(1-Y_s)-\theta_2Y_s\right)ds+\sqrt{Y_s(1-Y_s)}dW_s,\qquad
Y_0=y_0,\;\; \theta_1,\theta_2>0.
\end{equation}
For inference purposes, we consider the following reparametrisation:
$$\ds\gamma_1=\theta_1+\theta_2\quad \textrm{and} \quad \ds\gamma_2=
\frac{\theta_1}{\theta_1+\theta_2}.$$ The two new parameters
$\gamma_2$ and $\gamma_1$ represent the process' reversible mean and
the drift force towards it, respectively, and are expected to have low
posterior correlation. This is useful since it makes an uncorrelated
random walk proposal for $(\gamma_1,\gamma_2)$ a reasonable
choice. After this reparametrisation, the Lamperti transform leads to
a diffusion with unit diffusion coefficient and drift of the form
\[
\alpha(u;\theta)=
\frac{1}{2\sin(u)}\Big( \gamma_1 (2 \gamma_2 -1)  + (\gamma_1 - 1)
\cos(u)\Big), \qquad  u\in (0, \pi).
\]
One can check that $(\alpha^2 + \alpha')$ is not uniformly bounded below in the
state space for some region of the parametric space, however it can be bounded conditionally once upper and lower bounds for the diffusion path
are obtained through the layered Brownian bridge construction discussed above.

\begin{table}[h]
\centering
\begin{tabular}{c|c|c|c}
  \hline
             & mean & s.d. & 95\% C.I. \\ \hline
  $\gamma_1$ & 7.649 & 0.729 & (6.502,8.895) \\ \hline
  $\gamma_2$ & 0.507 & 0.012 & (0.486,0.527) \\  \hline
\end{tabular}\caption{Posterior statistics for the parameters.}\label{t1}
\end{table}

\begin{figure}[h]
	\centering
	\includegraphics[scale=0.2]{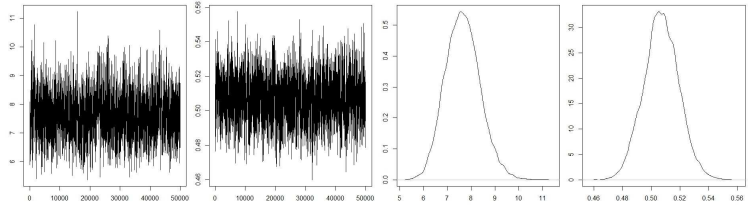}
	\caption{Posterior distribution of the parameter set: trace plots and marginal posterior densities.}\label{f1}
\end{figure}

We simulate 201 equally spaced observations in $[0,200]$ for $\gamma_1=8$ and $\gamma_2=0.5$. Uniform independent priors on the positive real line are adopted for $\gamma_1$ and $\gamma_2$. The chain runs for 50 thousand iterations with two consecutive updates for $\dot{X}$ at each iteration to improve the mixing of the chain. The proposal distribution for $(\gamma_1,\gamma_2)$ is a uniform random walk for each coordinate - $U(\gamma_1\pm0.65)$ and $U(\gamma_2\pm0.01)$. The acceptance rate of the Barker's step for the parameter vector was 0.357. The estimated posterior correlation of $\gamma_1$ and $\gamma_2$ was -0.005. Results are presented in Figure \ref{f1} and Table \ref{t1}. A burn-in of 2 thousand iterations is used to compute the posterior estimates.

\section{Acknowledgements}
This research has been supported by the Brazil Partnership Fund 2014/15. Krzysztof {\L}atuszy\'nski is supported by the Royal Society through
the  Royal Society University Research Fellowship. The first author would like to thank FAPEMIG for financial support.

\bibliographystyle{apalike}
\bibliography{barkers_revision}

\end{document}